\documentclass{article}
\usepackage{spconf,amsmath}
\usepackage{amsfonts} %
\usepackage{booktabs} %
\usepackage{multirow} %
\usepackage{tabularx}
\usepackage{diagbox} %
\usepackage[hidelinks,bookmarks=true,hypertexnames=true]{hyperref}

\newcommand{\tabincell}[2]{\begin{tabular}{@{}#1@{}}#2\end{tabular}}

\usepackage{cite}   %
\usepackage[pdftex]{graphicx}
\graphicspath{{./figs/}}
\DeclareGraphicsExtensions{.pdf,.jpeg,.png}

\usepackage{IEEEtrantools}

\title{Separating long-form speech with group-wise\\ permutation invariant training}
\name{ \begin{tabular}{c} Wangyou Zhang$^{1,2,*}$, Zhuo Chen$^{2}$, Naoyuki Kanda$^{2}$, Shujie Liu$^{2}$, Jinyu Li$^{2}$, Sefik Emre Eskimez$^{2}$,\\ Takuya Yoshioka$^{2}$, Xiong Xiao$^{2}$, Zhong Meng$^{2}$, Yanmin Qian$^{1}$, Furu Wei$^{2}$\end{tabular}\thanks{$^{*}$Work done during internship at Microsoft.}}
\address{
$^1$ MoE Key Lab of AI, X-LANCE Lab, CSE Dept, Shanghai Jiao Tong University, Shanghai China\\
$^2$ Microsoft Corporation}
\begin{document}
\ninept
\maketitle

\begin{abstract}
Multi-talker conversational speech processing has drawn many interests for various applications
such as meeting transcription. %
Speech separation is often required to handle overlapped speech that is commonly observed in conversation.
Although the original utterance-level permutation invariant training-based continuous speech separation approach has proven to be effective in various conditions, it lacks the ability to leverage the long-span relationship of utterances and is computationally inefficient due to the highly overlapped sliding windows.
To overcome these drawbacks, we propose a novel training scheme named Group-PIT, which allows direct training of the speech separation models on the long-form speech with a low computational cost for label assignment.
Two different speech separation approaches with Group-PIT are explored, including direct long-span speech separation and short-span speech separation with long-span tracking.
The experiments on the simulated meeting-style data demonstrate the effectiveness of our proposed approaches, especially in dealing with a very long speech input.

\end{abstract}


\vspace{-5pt}
\section{Introduction}
\label{sec:intro}
\vspace{-5pt}

Speech processing for multi-talker conversational speech, such as meeting recordings, is very challenging in the real world.
It differs from single-talker scenarios in two main aspects.
Firstly, it naturally contains overlapped speech from multiple speakers, so a speech separation process is often required.
Secondly, a conversation can be of any length without any segmentation, which poses a challenge to the long-form speech processing capability of the system.
There have been increasing interests in the conversational speech processing, including automatic speech recognition (ASR)~\cite{watanabe20b_chime,chang2021hypothesis,kanda2021comparative}, speech separation~\cite{yoshioka2018recognizing,chen2020continuous,li2021dual,wang2021localization}, and speaker diarization~\cite{von2019all,xiao2021microsoft}.
In this paper, we specifically focus on the speech separation problem for long-form speech. 

Continuous speech separation (CSS)~\cite{chen2020continuous} is a framework to convert long-form unsegmented audio into $N$ overlap-free audio streams.
In its representative instantiation with utterance-level permutation invariant training (uPIT)~\cite{yu2017permutation,kolbaek2017multitalker},
the input speech is first segmented by using a sliding window with overlaps, and speech separation is independently performed on each segment to generate $N$ separated signals.
The separated signals in adjacent segments are then aligned via a stitching algorithm.
This approach has not only proven to be effective in speech separation of simulated long-form signals~\cite{chen2021continuous,CSS_with_EETransformer}, but also shown large improvement in ASR~\cite{raj2021integration,yoshioka2019advances} and speaker diarization~\cite{xiao2021microsoft} tasks in realistic conversation scenarios.
However, there are some drawbacks in such a uPIT-CSS approach:
(1) It is computationally inefficient due to the large overlap between adjacent windows, which is essential for better stitching performance.
(2) More importantly, the uPIT-CSS approach can only model the short-span relationship of utterances, e.g.~1.6s in \cite{yoshioka2018recognizing}, as it assumes at most $N$ active speakers in each window where $N$ is typically 2.  When a long window is used for local separation, the assumption above is likely to be broken as more than $N$ speakers are likely to be present within a window. Therefore, its performance is limited due to the lack of access to a long-span context.

A recent study~\cite{neumann21_interspeech,von2021speeding} proposed a novel method tackling the above problems in the CSS framework, where
the authors show that the label assignment in long-form speech separation can be regarded as a graph coloring problem, which leads to a generalized uPIT criterion named Graph-PIT.
The computational complexity in the initial work~\cite{neumann21_interspeech} scales exponentially with the number of utterances in each segment, and was later reduced to be linear in the number of utterances via dynamic programming~\cite{von2021speeding}.

In this paper, %
we aim to solve the long-span speech processing approach without changing the PIT objective function.
We propose Group-PIT (gPIT in short), a simple training data construction strategy to address this problem, to allow the separation network
to directly process long-form speech in both training and inference stages.
We show that by carefully designing the data simulation procedure and arranging the long-form reference signal into utterance groups, the number of possible permutations in each long-form audio (e.g.~60s) can be constrained to $N!$
regardless of the number of active speakers and utterances. 
This allows training of speech separation models directly on long-form speech with the same training objective as uPIT, except that it is used for utterance groups rather than individual utterances.
We also explore different long-form speech processing approaches with Group-PIT.
Firstly, we show that the straightforward extension of CSS to gPIT-CSS with long-span separation can better process the long-form speech, which benefits from the direct long-span modeling.
Secondly, we explore a two-stage gPIT-CSS approach with short-span separation and long-span tracking.
This approach combines the properties of the local- and long-span processing, which is suitable for conditions where the long-form training data is difficult to obtain or simulate, e.g., realistic long-form conversation speech with spontaneous speaker interactions.
The effectiveness of our proposed methods is validated on the simulated meetings based on the WSJ corpus~\cite{wsj0,wsj1}.

\section{Stitching-based uPIT-CSS}
\label{sec:bg}
\vspace{-5pt}

\begin{figure}[t]
  \centering
  \includegraphics[width=\columnwidth]{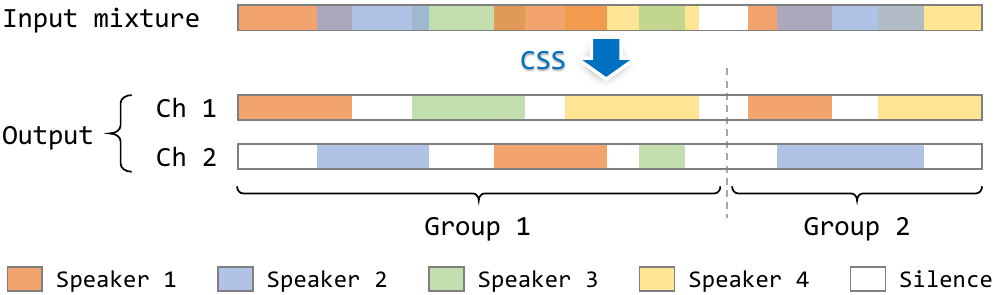}
\vspace{-3mm}
  \caption{The ideal CSS output channels ($N=2$). Blocks in different colors represent utterances from different speakers.}
  \label{fig:css_output}
\end{figure}

We suppose the long-form input speech mixture $\mathbf{Y}$ consists of $U$ utterances and in total $C$ speakers.
In the CSS framework, it is assumed that at most $N$ speakers are active at the same time so that all utterances in $\mathbf{Y}$ can be separated and placed into $N$ channels.
Each output channel has the same length as the input, and only contains overlap-free utterances, as shown in Fig.~\ref{fig:css_output}.

\begin{figure}[t]
  \centering
  \includegraphics[width=0.85\columnwidth]{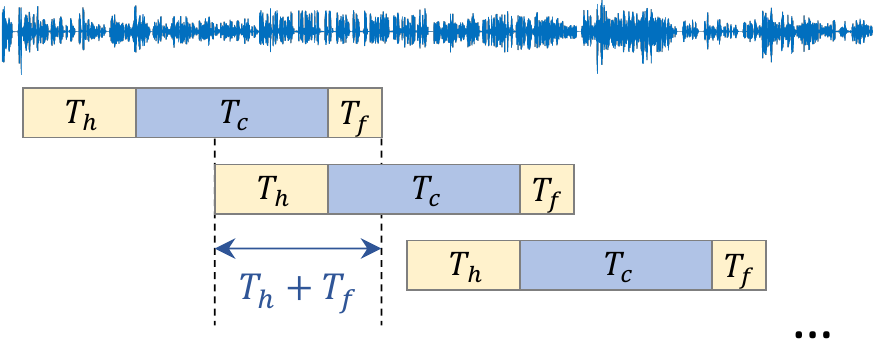}
  \vspace{-5mm}
  \caption{Segment-wise processing in stitching-based CSS. Separation is performed in each segment independently with a sliding window.}
  \label{fig:css}
\end{figure}

A typical CSS pipeline with uPIT-based speech separation ~\cite{yoshioka2018recognizing} is composed of three stages: segmentation, separation, and stitching.
As illustrated in Fig.~\ref{fig:css}, the segmentation stage divides the long-form audio into several overlapped segments using a fixed-length sliding window.
Each sliding window consists of three parts with $T_h$, $T_c$, and $T_f$ frames %
that represent the history, current, and future frames, respectively.
The overlap length between adjacent segments is $T_h+T_f$.
The speech separation is then performed on each segment independently to generate $N$ overlap-free signals.
Finally, the separation outputs in all segments are merged via a stitching algorithm to obtain the meeting-level separation result.
This is done by first finding the best permutation of output channels with the highest overall similarity in each pair of adjacent segments and permuting them accordingly.
Then, an overlap-and-average operation is performed along each channel across all segments.

It should be noted that the uPIT-based CSS assumes that the window length is small enough to only contain at most $N$ speakers so that uPIT-based speech separation models can be trained.
It makes the uPIT-CSS difficult to use a long window where more than $N$ speakers will be likely to appear
and potentially limits the modeling capacity due to the lack of access to the long-span context.

\vspace{-8pt}
\section{Group-PIT for long-span modeling}
\label{sec:long_span_pit}
\vspace{-5pt}
In this section, we introduce the proposed Group-PIT approach for long-form speech separation. First, we define the data arrangement in our proposed approach. Later, we propose two different approaches based on Group-PIT.

\vspace{-10pt}
\subsection{Group-PIT and the corresponding data arrangement}
\label{ssec:data}
\vspace{-3pt}

As illustrated in Fig.~\ref{fig:css_output}, in the original CSS pipeline, the placement of some separated utterances in different output channels may not be unique.
For example, the last two utterances (orange and yellow) in \texttt{Ch~\!1} can be swapped with the last utterance (blue) in \texttt{Ch~\!2}, while still satisfying the CSS constraint introduced in Section~\ref{sec:bg}.
This phenomenon is common and can easily happen when a relatively long silence exists in the midst of the input speech mixture.
As a result, the number of possible permutations of $U$ separated utterances in $N$ output channels can be up to $N^U$.
More specifically, if we define an \emph{utterance group} as a consecutive segment in which all utterances in $N$ channels only have $N!$ possible permutations, then the number of possible permutations in the CSS problem is up to $(N!)^G$, where $G$ is the number of utterance groups in the CSS output. 
In Fig.~\ref{fig:css_output}, we can see that $G=2$ and $N=2$, so there are $(2!)^2=4$ permutations in total.
Since this number increases exponentially with the number of utterance groups, it would be computationally expensive to extend CSS by using a longer window directly. 

To remedy this issue, we propose to proactively arrange the training data 
so that $G=1$ is guaranteed for every long-form sample.
This makes the number of possible permutations significantly small.
In the following discussion, we adopt $N=2$ as three-fold overlaps are rarely observed in real meetings~\cite{chen2020continuous}.
When simulating a long-form speech sample, we first generate two overlap-free reference signals corresponding to two output channels by iteratively appending $U$ utterances to either of the channels based on the following rules:
(i) the first utterance is appended to \texttt{Ch~\!1}, (ii) the $u$-th utterance is appended to the channel $n_u\in\{1,2\}$ where the end time, $e_{n_u}$, of the lastly-appended utterance is earlier than the end time, $e_{\bar{n_u}}$, of the lastly-appended utterance on the other channel. When the $u$-th utterance is appended, %
its onset is randomly sampled from the uniform distribution 
$\mathcal{U}(e_{n_u}, e_{\bar{n_u}})$.
After generating two overlap-free reference signals, they are mixed to form the long-form audio mixture for training.
This constraint guarantees that only one utterance group exists in each long-form speech sample.
Given the small number of possible permutations, 
we can apply the conventional uPIT criterion
except that it is applied for utterance groups rather than individual utterances.
We call this method Group-PIT.

Compared to our proposed method, Graph-PIT~\cite{neumann21_interspeech,von2021speeding} is a more generalized approach that directly extends uPIT for long-span modeling. 
On the other hand, our proposed method simplifies the permutation problem, thus reducing the computational cost during training. The proposed method can be viewed as a special solution, which leverages the prior knowledge in the CSS problem, of the Graph-PIT. Such prior knowledge can result in different behaviors for the separation network when $C>2$, while we focus on the case of $C=2$ in this paper and leave such cases for future work. It should be noted that with additional constraints, it is also possible for Graph-PIT to converge to the same solution as the proposed method.

Note that modeling long-form samples containing multiple utterance groups ($G>1$) during training is potentially inefficient.
Because from the practical perspective, it is relatively easy to detect long silence regions by applying voice activity detection (VAD) as a preprocessing. 
The input mixture can then be divided into chunks without such silence\cite{von2021speeding}, and the separation output for each chunk can still be regarded as a single utterance group. Note that in this work, utterances with short silence in between are considered to belong to the same utterance group.

\begin{figure*}[t]
  \centering
  \includegraphics[width=0.85\textwidth]{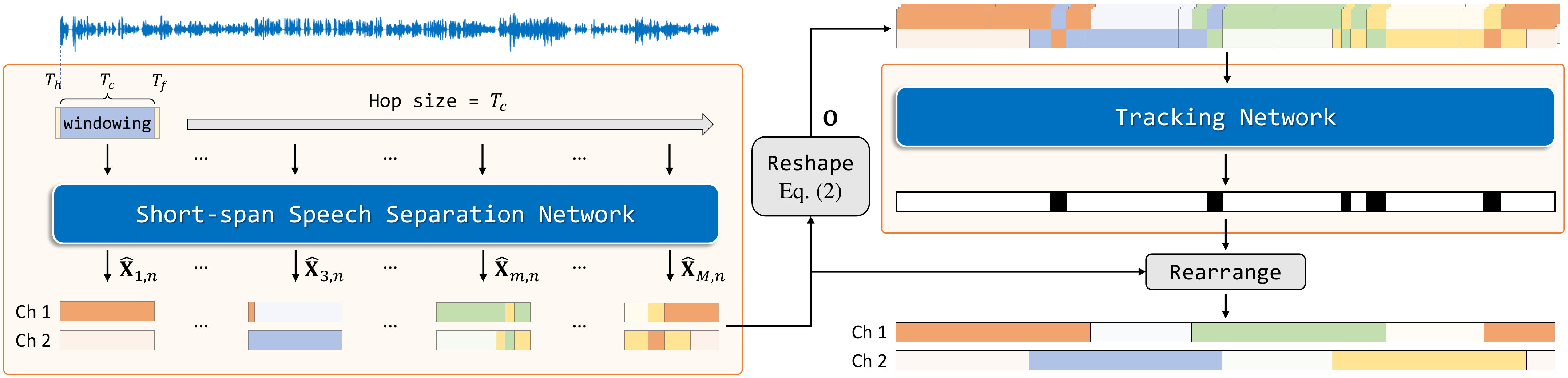}
  \caption{Proposed gPIT-CSS approach with short-span separation and long-span tracking.}
  \label{fig:sep_track}
  \vspace{-1.0em}
\end{figure*}
\vspace{-5pt}

\vspace{-5pt}
\subsection{gPIT-CSS with long-span separation}
\label{ssec:sess_pit}
\vspace{-2pt}

One straightforward way to extend the uPIT-CSS with $N=2$ by Group-PIT is using a longer sliding window that covers more than two utterances.
In the training stage, we assume that the reference signal $\mathbf{X}_n$ ($n=1,2$) only contains one utterance group.
The training objective is then given by:
{\setlength\abovedisplayskip{3pt plus 3pt minus 3pt}
\setlength\belowdisplayskip{3pt plus 3pt minus 3pt}
\begin{align}
    \mathcal{L}^{\text{(Group-PIT)}} = \min_{\pi \in \mathcal{P}_2} \sum_{n=1}^2 \mathcal{L}\big(\mathbf{X}_n, \hat{\mathbf{X}}_{\pi(n)}\big)\,,\label{eq:group_pit}
\end{align}
}where $\hat{\mathbf{X}}_{\pi(n)}$ is the $\pi(n)$-th output signal from the speech separation model, $\pi \in \mathcal{P}_2$ enumerates all possible permutations for $N=2$ channels, and $\pi(n)$ denotes the permuted index for the $n$-th channel.
$\mathcal{L}$ is the loss function either in the time domain or frequency domain.

In the inference stage, the same stitching-based process as in Section~\ref{sec:bg} is used for processing the entire meeting, except that a much longer window size can be used.
It is thus possible to directly utilize the long-span audio context for better speech separation.

\vspace{-8pt}
\subsection{\protect\resizebox{0.92\columnwidth}{!}{gPIT-CSS with short-span separation and long-span tracking}}
\label{ssec:sep_track}
\vspace{-3pt}

The separation approach in Section~\ref{ssec:sess_pit} solves the long-span separation problem in one shot. However, it usually requires matched training data to maximize its advantage in long-form modeling, which is not always available in practice.
For example, it is challenging to simulate all the varieties in realistic long-form conversation speech that includes spontaneous speaker interactions.
Without matched data, the long-span modeling could be potentially sub-optimal.
Therefore, we explore another approach to apply Group-PIT in the CSS pipeline, where the speech separation procedure of one long-segment is decomposed to short-span separation and long-span tracking procedures. 

The overview to process one long audio segment (such as 24s) is depicted in Fig.~\ref{fig:sep_track}.
In this approach, the long audio segment is further segmented by short sliding windows (such as 4s) 
with almost no overlap where $T_h=T_f=1$ frame. 
For each short window indexed by $m$, a short-span separation model trained with the conventional uPIT objective function is applied
to generate two overlap-free signals $\hat{\mathbf{X}}_{m,n}$ where $n\in\{1,2\}$.
The long-span tracking network is then applied on the  separated signals from all short windows 
to predict the frame-wise permutation for output channels as following:
{\setlength\abovedisplayskip{3pt plus 3pt minus 3pt}
\setlength\belowdisplayskip{3pt plus 3pt minus 3pt}
\begin{align}
    &\!\!\mathbf{O}\!=\!\operatorname{Concat}_{m=1}^M(\operatorname{Splice}([\hat{\mathbf{X}}_{m,1}^{\mathsf{T}};\hat{\mathbf{X}}_{m,2}^{\mathsf{T}}]^{\mathsf{T}}))\,,\label{eq:track_feat}\\
    &\!\!\mathbf{P} = \operatorname{TrackNet}(\mathbf{O})\,.\label{eq:track_out}
\end{align}
}Here,
$\operatorname{Splice}(\cdot)$ denotes the operation that stacks each frame and its adjacent frames along the feature dimension.
$\operatorname{Concat}_{m=1}^M(\cdot)$ denotes the operation to concatenate all features from $M$ short windows 
along the frame dimension.
The term $\mathbf{P} \in \mathbb{R}^{1\times T}$ represents the frame-wise permutation indicator, where ``1'' indicates swapping the two separation results from current frame in \texttt{Ch 1} and \texttt{Ch 2}, while ``0'' means no change. $T$ refers to the length of the input sequence. According to the permutation indicator, 
the short-span separation result $\hat{\mathbf{X}}_{m,n}$ is rearranged to form the final long-span output signal for
each channel.
Note that the above procedure is the explanation for the speech separation of one (relatively long) audio segment,
which is still shorter than the duration of an entire recording.
The entire recording is processed with the stitching algorithm as used in the gPIT-CSS with long-span separation
in Section~\ref{ssec:sess_pit}.

The cross-entropy loss is used to train the tracking network:
{\setlength\abovedisplayskip{3pt plus 3pt minus 3pt}
\setlength\belowdisplayskip{3pt plus 3pt minus 3pt}
\begin{align}
    \mathcal{L}_{\text{track}} &= \operatorname{CrossEntropy}(\mathbf{P}, \mathbf{R})\,,\label{eq:track_loss}
\end{align}
}where $\mathbf{R} \in \mathbb{R}^{1 \times T}$ is the oracle frame-wise permutation.
The oracle permutation label is formed by comparing the frame-wise separation result with the two-channel reference spectrum. 
The data arrangement from Group-PIT is applied to construct the reference signals, i.e., after tracking alignment, the final separation result should follow the CSS arrangement as shown in Fig. \ref{fig:css_output}. 
We freeze the short-span separation model when training the tracking network.

Note that the idea of combining short-span separation and long-span tracking was already investigated in the literature~\cite{liu2019divide}, but only for the utterance-level mixtures.
Therefore, it is still unclear how well this approach works in the CSS framework.
Our proposed extension with Group-PIT naturally fills this gap. 

\vspace{-8pt}
\section{Experimental Setup}
\label{sec:exp}
\vspace{-5pt}
\subsection{Data description}
\label{ssec:exp_data}

We experimented with simulated multi-talker recordings based on the WSJ corpus~\cite{wsj0,wsj1}.
The training and development sets were simulated based on the WSJ1~\cite{wsj1} training set, with 283 speakers in total.
The evaluation set was simulated based on the \texttt{si\_dt\_05} and \texttt{si\_et\_05} subsets from WSJ0~\cite{wsj0}, with 18 speakers in total.
The sampling rate of the audio was 16kHz. %
The simulation of all datasets follows the description in Section~\ref{ssec:data}.
The number of speakers ranges from 2 to 5, while the meeting length is fixed to 80s.
For all datasets, we simulated two types of mixtures, i.e.~partially overlapped mixtures (\texttt{partial}) and sequential mixtures (\texttt{seq.}).
For training, development and evaluation, we used 27000, 2992, 2999 overlapped samples and 8000, 1500, 3000 sequential samples, respectively.
For overlapped mixtures, the overlap ratio ranges from 20\% to 60\%.
For sequential mixtures, note that we considered sequential utterances with a short pause ($<$0.5s) belong to the same utterance group, and constrain them to be assigned to different channels even they are not overlapped. This property of separation networks has been shown to be important to handle quick speaker turns in real conversation~\cite{yoshioka2019advances}.

\vspace{-8pt}
\subsection{Network Architectures}
\label{ssec:exp_net}
\vspace{-2pt}
We adopt the time-frequency masking~\cite{Supervised-Wang2018} based speech separation method to examine the effectiveness of the proposed approaches.
The window size and hop size for short-time Fourier transform (STFT) are 512 and 256, respectively.
The loss function $\mathcal{L}$ in Eq.~(\ref{eq:group_pit}) is the L2 loss between estimated and reference magnitude spectra.
For the gPIT-CSS approach with long-span separation, we adopt the dual-path transformer (DP-transformer)~\cite{Dual_path-Luo2020,Attention-Subakan2021} architecture for its capability and efficiency in long sequence modeling.
It consists of 16 encoder layers with 4 attention heads, and each layer has 128 attention dimensions and 1024 FF dimensions.
For the gPIT-CSS approach with separation and tracking, we adopt the transformer model with 16 layers and a similar amount of parameters for short-span separation.
For the tracking network, we also adopt the DP-transformer architecture for long-span modeling, which consists of 16 encoder layers with 128 attention dimensions and 1024 FF dimensions.
The chunk size and hop size in the inter- and intra-chunk processing in all DP-transformer models are 150 and 75, respectively.
The batch size is 96 for training Group-PIT models with a 4s sliding window on 8 GPUs.
For other window lengths, we adjust the batch size accordingly to fit approximately the same amount of data into each batch as long as the memory can hold.
The AdamW optimizer is used for training.

\vspace{-10pt}
\section{Experimental Results}
\label{sec:exp_result}
\vspace{-9pt}

\begin{table}[t]
    \caption{Average utterance-level SI-SNR (dB) of gPIT-CSS based long-span separation models with different sliding window sizes.}
    \label{tab:gpit_dpt}
    \centering
    \resizebox{\columnwidth}{!}{%
    \begin{tabular}{lc|cccc}
    \toprule
    \multirow{2}{*}{Model} & \multirow{2}{*}{$T_{\text{tr}}$ (s)} & \multicolumn{4}{c}{Sliding window size (s)} \\
    & & 4 & 16 & 32 & 60 \\
    \midrule
    Original \texttt{partial} mixture & - & \multicolumn{4}{c}{--------------- 2.84 ---------------} \\
    \hline
    \multirow{4}{*}{\tabincell{l}{gPIT-CSS with long-span\\separation}} & 4 & 9.02 & 2.70 & 2.40 & 2.31 \\
    & 16 & 0.94 & 13.92 & 7.08 & 6.46 \\
    & 32 & 1.04 & 13.47 & 14.10 & 12.26 \\
    & 60 & 1.84 & 11.61 & 13.01 & \textbf{14.58} \\
    \hline
    \multirow{4}{*}{$\ \ $+ Oracle permutation} & 4 & 11.93 & 8.59 & 5.59 & 3.66 \\
    & 16 & 8.55 & 15.64 & 11.18 & 8.19 \\
    & 32 & 8.84 & 15.87 & \textbf{16.12} & 13.67 \\
    & 60 & 8.42 & 15.00 & 15.38 & 15.25 \\
    \bottomrule
    \end{tabular}%
    }
\end{table}

\begin{table}[t]
    \vspace{-5pt}
    \caption{Average frame-wise accuracy (\%) of gPIT-CSS based long-span separation models with different sliding window sizes.}
    \label{tab:gpit_dpt_seq}
    \centering
    \resizebox{\columnwidth}{!}{%
    \begin{tabular}{lc|cccc}
    \toprule
    \multirow{2}{*}{Model} & \multirow{2}{*}{$T_{\text{tr}}$ (s)} & \multicolumn{4}{c}{Sliding window size (s)} \\
    & & 4 & 16 & 32 & 60 \\
    \midrule
    Original \texttt{seq.} mixture & - & \multicolumn{4}{c}{--------------- 2.84 ---------------} \\
    \hline
    \multirow{4}{*}{\tabincell{l}{gPIT-CSS with long-span\\separation}} & 4 & 77.87 & 63.12 & 62.71 & 61.99 \\
    & 16 & 56.65 & 94.22 & 77.65 & 77.00 \\
    & 32 & 53.46 & 94.21 & 94.66 & 91.92 \\
    & 60 & 53.53 & 90.43 & 93.40 & \textbf{97.45} \\
    \bottomrule
    \end{tabular}%
    }
\end{table}

\subsection{gPIT-CSS with long-span separation}
\label{ssec:stitching}
\vspace{-3pt}
As mentioned in Section~\ref{ssec:sess_pit}, the proposed Group-PIT allows training of speech separation models on much longer segments than uPIT.
Therefore, we first compared the performance of direct long-span separation models trained with different window lengths $T_{\text{tr}}$.
The best permutation of the meeting-level separation output channels is first determined, and the oracle utterance boundaries in each channel are then used to calculate the utterance-level scale-invariant signal-to-noise ratio (SI-SNR)~\cite{leroux2019sdr}.
The overlap between adjacent windows is set to 2s by default, i.e. $T_h = T_f = 1s$ for all models.

Table~\ref{tab:gpit_dpt} shows the separation performance on the overlapped evaluation data (\texttt{partial}).
It is shown that when evaluated with different sliding window lengths, models trained with a longer window tend to have better performance.
This verifies our conjecture that a longer context can benefit the separation of long-form audios.
In all conditions, the best performance is achieved when the same window length is used for both training and evaluation. 
In addition, we can observe that models trained with longer windows tend to reach the performance with oracle permutations\footnote{Here, ``oracle permutations'' means
using the reference signal to determine the permutation of each window to stitch adjacent separated segments.},
which further demonstrates the effectiveness of the proposed approach. Note that the setting of 4s training data is usually adopted by uPIT-CSS systems, and the gPIT-CSS with 4s can serve as a reference for uPIT-CSS. However, it should be noted that gPIT and uPIT are not equivalent for this condition, as the gPIT training data might contain more than 2 speakers in one 4s training sample while uPIT strictly requires no more than 2 speakers for each sample.

For the sequential evaluation data (\texttt{seq.}), since no overlap exists, the SI-SNRs of the separation outputs tend to be very large ($>30$dB), which is inappropriate to compare due to the nonlinear scale in SI-SNR.
Instead, we compare the frame-wise accuracy of speaker assignment in each output channel in Table~\ref{tab:gpit_dpt_seq}.
This is obtained by calculating the percentage of speaker turns in the best frame-wise permutation based on the final meeting-level separation output.
It can be seen that models trained with longer windows also show higher frame-wise accuracies on the sequential mixture, which further shows the benefit of the proposed approach.

\begin{table}[t]
    \caption{Performance comparison of two different gPIT-CSS approaches. ``stitching'' denotes the gPIT-CSS approach with long-span separation, while ``tracking'' denotes the gPIT-CSS approach with short-span separation and long-span tracking.}
    \label{tab:gpit_track}
    \centering
    \resizebox{\columnwidth}{!}{%
    \begin{tabular}{l|lcc}
    \toprule
    Model & Approach & Tracking acc. & SI-SNR \\
    \midrule
    Original \texttt{partial} mixture & no processing & - & 2.84 \\
    \hline
    \multirow{3}{*}{gPIT-CSS ($T_{\text{tr}}=2$s)} & + stitching & - & 4.36 \\
    & + tracking & 91.81\% & 9.21 \\
    & + oracle tracking & 100\% & 17.16 \\
    \hline
    \multirow{3}{*}{gPIT-CSS ($T_{\text{tr}}=4$s)} & + stitching & - & 7.50 \\
    & + tracking & 90.81\% & 8.22 \\
    & + oracle tracking & 100\% & 17.20 \\
    \bottomrule
    \end{tabular}%
    }
    \vspace{-9pt}
\end{table}

\vspace{-9pt}
\subsection{\protect\resizebox{0.92\columnwidth}{!}{gPIT-CSS with short-span separation and long-span tracking}}
\vspace{-5pt}
In this section, we evaluate the gPIT-CSS approach with short-span separation and long-span tracking.
In contrast with the relatively large window length and overlap length used in Section~\ref{ssec:sess_pit}, we only use a short sliding window (2s and 4s) with a 2-frame overlap for short-span separation.
The tracking network is trained and evaluated using a 24s sliding window.
The overlap between adjacent tracking windows is 12s and 2s for training and evaluation, respectively.
Table~\ref{tab:gpit_track} shows the performance of tracking-based models trained with different window lengths.
Although the frame-wise tracking accuracy is not low, the overall SI-SNR performance is not as good as the direct long-span separation approaches with the best stitching window configuration in Table~\ref{tab:gpit_dpt}, which suggests this approach is more sensitive to frame-wise tracking errors.
However, such comparison is unfair because much longer overlap sizes are used to achieve good performance in Table~\ref{tab:gpit_dpt}, leading to higher computational overhead.
If we reduce the overlap size to only 2 frames, as shown in Table~\ref{tab:gpit_track}, the performance of the long-span speech separation (denoted as ``stitching'' in Table~\ref{tab:gpit_track}) is severely degraded.
On the other hand, the tracking-based approach can significantly improve the final separation result while enjoying a much lower computational cost\footnote{The computational cost for our tracking network is roughly one third of the cost required for uPIT-CSS with 2s overlap, and the total computational cost becomes lower even with the overhead for the tracking network.}.
It is especially helpful when a shorter separation window is used, as more improvement is achieved with $T_{\text{tr}}$\,=\,2s over $T_{\text{tr}}$\,=\,4s.

\vspace{-8pt}
\section{Conclusion}
\label{sec:conclusion}
\vspace{-7pt}
In this paper, we explored the long-span speech separation approaches in the meeting scenario.
A novel training scheme called Group-PIT was proposed to cope with the permutation problem in long-form speech.
We showed that Group-PIT-based speech separation models can be trained directly on the arranged long-form speech with the same computational complexity as in uPIT.
Moreover, we explored two different Group-PIT-based speech separation approaches for long-span speech processing,
and their effectiveness was validated on the simulated data based on the WSJ corpus.

\bibliographystyle{IEEEtran}
\bibliography{Bibliography}

\end{document}